\documentstyle[psfig]{l-aa}
\begin{document}
\thesaurus{08              % A&A Section 6: Form. struct. and evolut. of stars
              (12.04.1;    % Dark matter
               12.07.1)}  % Gravitational Lensing;
\title{Analysis of the OGLE microlensing candidates using the image subtraction method.} 
\author{C. Alard \inst{1,2}}
\institute
{DASGAL,Observatoire de Paris, 61 Avenue de l'observatoire, F-75014
 Paris, France
\and
 IAP, 91bis boulevard Arago, F-75014 Paris, France.}
\offprints{C. Alard}
\date{Received ......; accepted ......}
\maketitle
\begin{abstract}
 The light curves of the OGLE microlensing candidates have been reconstructed
 using the image subtraction method. A large improvement of the
 photometric accuracy has been found in comparison with previous processing
 of the data with DoPHOT. On the mean, the residuals to the fit 
 of a microlensing
 light curve are improved by a factor of 2 for baseline data points, and by
 a factor of 2.5 during magnification. The largest improvement was found for 
 the OGLE \#5 event, where we get an accuracy 7.5 times better 
 than with DoPHOT. Despite some defects in the old CCD
 used during the OGLE I experiment
 we obtain most of the time errors that are only 30 \%
 to 40 \% in excess of the photon noise. Previous experiment showed that
 with modern
 CCD chips (OGLE II), residuals much closer to the photon noise were obtained.
 The better photometric quality enabled us
 to find a low amplitude, long term variability in the
 OGLE \#12 and OGLE \#11 baseline magnitude. We also found that the shape of
 the OGLE \#14 candidate light curve is fairly inconsistent with microlensing 
 of a point source by a point lens. A dramatic
 change in the light curve of the OGLE \#9 candidate was also found, 
 which indicates that very large biases can be present in data processed
 with DoPHOT. To conclude we made a detailed analysis of the blending issue.
 It is found that OGLE \#5 and OGLE \#6 are very likely highly blended
 microlensing events. These events result from the magnification 
 of a faint star that would have been undetectable without microlensing.
\end{abstract}
\section{INTRODUCTION}
 The undergoing microlensing surveys (OGLE, EROS, MACHO)
 are currently providing us with 
 an enormous amount of images of dense crowded stellar fields. This
 contribution of new data stimulated software development. Several
 successful implementations based on the DoPHOT software (Udalski {\it et al.}
 1993 Bennet {\it et al.} 1992) have been used so far to extract the stellar 
 magnitudes from the images. Other software, like PEIDA, which has 
 been developed by the EROS group
 (Ansari {\it et al.} 1996) has also been used with some success. 
 However, all these packages
 use the same basic profile fitting method to estimate the magnitudes of the 
 stars present in the image. This kind of method is not optimal in the case
 of microlensing surveys, as our aim is to detect and measure variations only.
 Purely differential methods, like the image subtraction technique should give
 better results. The recent release by the OGLE collaboration 
 (Wozniak \& Szymanski 1998) of a collection of small images centered on
 each OGLE microlensing candidate, and the corresponding light curves,
 offers a nice opportunity to test this
 assumption. We will analyze these images with the method recently
 developed by Alard \& Lupton (1998). This method is designed to give 
 optimal results, and it is well suited for small images, as all the pixels
 can be used to derived the best kernel solution. We will compare the image
 subtraction light curves with the light curves obtained by 
 Wozniak \& Szymanski (1998) using the DoPHOT software (Schechter, Mateo, \&
 Saha, 1993).
\section{The subtracted images.}
We have a few hundred 101x101 images for each OGLE candidate. 
%These images include defects most of the time. These defects can be bright
% saturated stars, but it seems that there are some other kind of defects
% due to the poor general quality of the CCD camera. 
 These OGLE I images show
 a mean quality that seems to be well below the quality of the OGLE II images
 (Udalski {\it et al.} 1997). We remind that a  subsample of the OGLE
  II images 
 has already been already processed with the image subtraction method (Alard
 \& Lupton 1998). 
\subsection{Image registration and re-sampling.}
 As it has been already described in Alard and Lupton (1998), we register
 the images by fitting the positions of the brightest stars. For these
 small images, we used a polynomial transform of order 1. Each image is then
 transformed to the astrometric system of a reference image by using
 a bicubic spline interpolation. This procedure provides us with a collection
 of images that have the same sampling but different seeings.
\subsection{Building a reference for image subtraction.}
 We found that better results are obtained if we could derive a high
 quality and almost noise free reference image. We approach this requirement
 by stacking between 10 and 20 of our best seeing (interpolated) images. Using
 this image we also compute a ``masked'' image in which all obvious defects
 like saturated stars, or bad CCD regions are masked by putting the relevant
 pixels in the reference frame to zero. The image subtraction 
 program will use this masked image to derive the
 list of pixels that should not be fitted in order 
 to derive the kernel solution.
\subsection{Image subtraction.}
As it has been already described in Alard \& Lupton (1998), the core of the
 problem is to seek for a convolution kernel that will transform our reference
 frame
 to exactly the same seeing as another frame. To find the best kernel
 solution we have to solve a set of linear least-square equations. We keep the
 same general kernel decomposition that we already used for the OGLE II data.
 However, this time we have to modify the code in order to take into account
 the defects in the reference image. If we did not take this precaution, the
 convolution with the kernel would spread the defects and pollute the fit.
 To solve this problem 
% we have to consider that
 the linear least square matrix
 is built by convolving the reference image with each of the vectors we use
 for the kernel decomposition. We do not wish to calculate the convolution at
 the pixels we have masked. But the problem will occur just at the border
 of the masked areas. To compute the convolution at a given position we need
 the values of the pixels all around this position, within some limiting radius.
 If some area around our position falls within the masked area we can
 not get any reasonable value for the relevant pixels. But we can find a 
 straightforward solution to this problem by not using these pixels to 
 calculate the convolution, and to compensate just by re-normalizing the 
 filter. 
 We adopt the sum of the filter to be one, and we re-normalize the filter
 so that its sum over all the pixels we have not discarded
 is again one. There are however some convolution filters that have a zero sum.
 We solve this difficulty by splitting the filter into 2 different filters
 that do not have a zero sum. We processed the frames for all the OGLE
 candidates with this method, and we found that the treatment of the defects
 was very satisfactory. 
\subsection{Quality of the subtracted images.}
 For some of the subtracted images we get residuals that are consistent with
 the Poisson fluctuations. However for a large number of frames we found
 some systematic residual patterns which seems to be related to the
 %a bizzare
 behavior of some of the CCD columns. It also appeared that the residuals
 associated with most of the bright stars were very satisfactory, showing
 that we reached a good kernel solution, but some showed
 %curious 
 systematics for one of the bright stars (somewhat in excess of the Poisson
 fluctuation). As a consequence of these defects the photon noise 
 limit which we easily reached with OGLE II data (Alard \& Lupton 1998) 
 is not as clearly reached in the case of these OGLE I data. The same
 image subtraction software has also been used for EROS data (Afonso 
 {\it et al.} 1998), and as in the case of OGLE II the residuals were
 in good agreement with the photon noise. 
 %We must conclude that the CCD
 %chip used in the OGLE I experiment was especially bad, and that such problems
 %should not be encountered any more with more recent CCD chips.
%
\section{Photometry of the subtracted images.}
 In the first paper concerning this image subtraction method, we used an easy
 and straightforward way to estimate the flux in each subtracted image. Our 
 aim was not to produce the best quality light curves, but just to show the 
 consistency with the photon noise. We 
 estimated the total flux by taking all the flux in a large aperture
 radius. This is of course not optimal. A better way to proceed is to make
 aperture photometry in a small radius only, and to compute an aperture 
 correction. It is also possible to use a generic profile fitting technique,
 by assuming that the profile is a bi-dimensional gaussian that we fit
 for each image. However, some experiments show that the result is not better
 than with corrected aperture photometry. It is also important to point out that
 in the case of profile fitting photometry small biases are introduced by the
 fact that the analytical model for the profile never matches perfectly the
 real PSF.
 We make a differential correction of this bias by comparing the flux of bright
 stars calculated by profile fitting in the image to the flux we estimated 
 in the reference image using the same method. The correction factor we derive 
 this way is about a few percent. To improve the photometry we need to derive
 a very accurate PSF model. This is not an easy task in very crowded fields.
\subsection{Optimized photometry of the subtracted images.}
This time we do not derive an analytical model for the PSF.
 Actually all we need to know is the PSF at the position of the star we
 want measure, i.e. the variable.
 We select a few bright stars in the
 image. Around each of these bright stars, we extract sub-images. 
 We also extract a sub-image around the star to measure. 
 We then re-center the bright stars, so that their exact position in their
 sub-images is the same as the position of the measured star in its sub-image.
 For this re-centering we use bicubic splines. 
 We emphasize that in order to get
 an unbiased position for the variable object (not the position of the nearest
 star), we stack all the subtracted images for which we find a significant 
 signal at the location of the candidate. This way we build a high signal to 
 noise image of the differential variations. We then use this image to estimate
 the center of the microlensing candidate.
 We also normalize
 each bright star sub-image by using a cross-product with a gaussian. We then
 do a median stacking of all sub-images, in order to get a numerical
 PSF bitmap. We use this PSF bitmap to estimate the flux in the relevant 
 subtracted image. Of course, as we have already explained, we
 systematically apply the differential fitting correction, in order to get
 rid of small discrepancies
 in our PSF model; although this time the discrepancy is very small. This 
 procedure gives the best results. We found a mean gain of about 20 to 30 \%
 in the accuracy of the photometry in comparison to the corrected aperture 
 photometry. It certainly indicates that most of
 the PSF's have very complicated shapes. 
\section{Fit of microlensing light curves to the data.}
 In order to make a comparison between the light curves obtained with
 DoPHOT (Wozniak \& Szymanski, 1998) and the image subtraction method, 
 we fitted theoretical microlensing
 light curves to both sets of data. In the fitted parameters we include the
 blending parameter, as blending of these faint sources is likely. To make
 a comparison as accurate as possible, we selected the data points that
 were both in the DoPHOT and the image subtraction light curves. Some images
 were defective and they were not in the DoPHOT light curve.
 In the case of image
 subtraction almost all the frames were treated. We could not estimate
 the flux with this method only in some rare cases (for instance when 
 a saturation spike 
 originating from a bright star came across the star image). We had
 no problem to process very bad seeings with the image subtraction method. 
\subsection{Analysis of the Chi-square and residuals to the fit.}
 To make comparison of DOPHOT versus image subtraction errors, we
 re-scaled each data set by dividing by the amplitude. We estimate this
 amplitude as the difference between the base line value and the value
 of the fitted curve at the location of the data point that shows the
 maximum magnification. We have also analyzed the consistency of the image
 subtraction errors with the Poisson noise. For this particular task we
 calculated the Chi-square of the distribution, by summing the square
 of the errors weighted with the Poisson (or photon noise) deviation.
 The Poisson deviation is estimated by calculating the variance
 of our magnitude estimator. In this calculation we
 assume that the Poisson distribution
 is well approximated by a Gaussian distribution with 
 $\sigma=\sqrt(counts/adu\_to\_el)$ ($adu\_to\_el$ is the conversion from
 ADU to electrons).  
 The results obtained are summarized in table 1. 
 We see that the mean improvement we get with
 the image subtraction method is about a factor of 2 in comparison to DoPHOT.
 It seems we gain even a little bit more if we compare the residuals during
 the magnification. Thus it is clear that a dramatic improvement is 
 performed by using the image subtraction method. Additionally, most
 of the time we get errors that are only 10 \% to 40 \% in excess of the photon
 noise expectation. Some candidate events have errors considerably larger 
 than the rest, and they introduce some biase in the 
 mean value of the improvement;
 we will discuss this issue later. 

 To illustrate the improvement
 that we get by using the image subtraction method we present 3 different
 cases. First we present an event with mean improvement, for which the 
 residuals are smaller by a factor of 2 (OGLE \#1, Fig. 1). 
 In the second figure we
 present the data for a case of a modest improvement (OGLE \#2), the residuals
 are smaller by only 37 \% for the baseline and by 87 \% during the 
 magnification.
 However, looking at the light curve, we see that the improvement is very
 significant. In particular, there are
 2 bad seeing points at the top of the light
 curve that deviate a lot in the DoPHOT, but have much smaller errors
 in the image subtraction light curve. This does illustrates very well the 
 ability of the method to deal with large seeing variations. Concerning the
 residuals along the baseline, it is important to note that at least one point
 has a very large deviation, and it is not consistent with Poisson
 statistics. We must emphasize again that the CCD used during the OGLE I 
 project
 had lot of defects and that might contribute to the errors. 

 To conclude we present the examples of a large improvement (OGLE \#4), and
 a very large improvement (OGLE \#5, Fig. 3). In the case of OGLE \#4 the
 residuals
 are reduced by a factor 2.65 during the baseline and 
 3.4 during the magnification. It is
 important to note that this result has been obtained despite especially
 bad image quality for this event, and yet we get a Chi-square
 that is very close to the photon noise expectation. For OGLE \#5, 
 the residuals are reduced by a factor 7.5 during the magnification.
 This shows than the image subtraction method can lead to very
 dramatic improvement. This huge improvement can probably be explained 
 by the fact that OGLE \#5 is a case of unresolved (highly blended) source
 (Alard 1997). 
\subsection{Light curves with large deviations from photon noise.}
The largest value of the Chi-square is for the OGLE \#3 event. It is
 good to note that OGLE \#3 is also the brightest of the candidates, and
 that it might indicate that with the OGLE I data we have some trouble
 in reaching the photon noise limit for bright stars. 
 Considering that a typical photometric error is only $ \sim 0.5\% $,
 any errors in the flat fielding may contribute
 to the global error budget . 
 It may also be related
 to the quality of the CCD chip, and in particular to the stability of
 the PSF to the level of 1 \%.
 The analysis
 of subtracted images shows residuals with different shapes even
 for the bright stars that are quite close to each other. These
 differences can not be
 explained by the seeing variations as the exposure time (10 min) is
 relatively long. We did not 
 found such problems in our analysis
 of OGLE II data (Alard \& Lupton 1998). To conclude the discussion
 on OGLE \#3 it is important to emphasize that even if the image subtraction
 is a bit far from the photon noise deviation expected from our profile
 fitting method, it is about 2 times better than DoPHOT. \\\\
 There are some faint stars that show large baseline residuals. These are:
 OGLE \#2, OGLE, \#11, OGLE \#12, OGLE \#14. In the case of OGLE \#2 we get
 a bad Chi-square for both baseline and magnification data. The case of 
 OGLE \#11 and OGLE \#12 is different, since Chi-square is bad during
 the baseline only. Finally, OGLE \#14 is the most peculiar since it shows
 a very bad Chi-square during the magnification only. \\\\
 It was not possible to find any reason for the bad Chi-square obtained for
 OGLE \#2. The residuals do not show any systematics, and no periodic signal
 could be identified. It appears that OGLE \#2 is quite different from
 the other OGLE events, since it lies close to the edge of the CCD. Considering
 the poor quality of this CCD this situation might lead to additional
 errors, especially in the flat fields. \\\\
 The cases of OGLE \#11 and OGLE \#12
 appear to be completely different since we can observe systematic patterns
 in the residuals (Fig. 4 and 5.). In the case of OGLE \#12, it is
 clear that a systematic trend of the baseline magnitude is observed. 
 A similar trend is also visible in the DoPHOT light curve, although it
 is less clear; it is also less reliable, as the image
 subtraction is purely differential and free of systematics. For OGLE \#11
 we also observe a systematic deviation of a clump of points near the date 1200.
 It seems to indicate that these 2 candidates have long term variations which
 resemble the variations observed for the OGLE \#10 
 candidate (cf. Fig. 6). \\\\
 The OGLE \#14 candidate does not show too much
 deviations during the baseline, but it presents very significant deviations
 from the
 best fit during the magnification. In particular, a group of points near the
 tip of the descending branch shows large systematic deviations. There is also
 a point which shows large deviations both, in the image subtraction and DoPHOT
 light curves. However the image corresponding to this data point does not
 show any particular defect, and seems to have a general good quality. We
 conclude that the most likely explanation is that OGLE \#14 could be a
 variable star, possibly a cataclysmic star, since we
 do not have a point on the rising branch that could discard this type of a
 variable. Another possibility would be that OGLE \#14 be a case of
 lensing by a binary source, or lensing of a binary source. However
 with the current data set it seems difficult to study this possibility.\\ \\
 These 4 microlensing candidates are also those which show the smallest
 improvements with the image subtraction method. This is not surprising,
 since it seems that in this case there is some additional error or 
 intrinsic variability that
 the image subtraction can't improve. There is another case where
 the improvement due to image subtraction is rather small, this is the 
 candidate OGLE \#9. However, the DoPHOT light curve and the
 image subtraction light curve are very different (see Fig. 8), and
 we wonder if the DoPHOT and the image subtraction measured exactly the
 same object. Looking at the composite reference image, we see that in
 the region occupied by OGLE \#9 there are many very faint unresolved objects,
 and that some confusion is likely. Sometimes DoPHOT may find 1 or 2 objects,
 depending on the seeing. \\\\
 To conclude this discussion, we present the mean of the residual ratios
 and the Chi-square when the 4 previous OGLE candidates are rejected. As it
 seems that for these 4 candidates additional (uncontrolled) errors or
 variations are
 present, it is difficult to derive a reliable statistics from them
 (see table 1).
\section{Blending analysis}
 In these crowded fields blending of the source with fainter stars is
 likely (classical blending), and this possibility needs 
 to be investigated. However another
 kind of blending is also possible: the source is very faint, it is unresolved
 and it is seen only during the magnification (Alard 1997, Han 1997). This kind
 of microlensing events is likely to give an important contribution to the
 optical depth, and it is important to try to detect them. The task
 should be easer than in the case of classical blending, as the blending
 of the source is very large and it should make a more noticeable modifications
 of the light curve. To investigate the blending issue we need to get the
 baseline magnitude of the star associated with the variations. For this
 particular task, we used the median PSF model that we derived previously
 to fit the magnitude of the star on the reference image. For this PSF
 fitting we used the coordinates that were found for the variable
 by stacking the subtracted images. 
 \subsection{Fit of models with and without blending to the light curves.}
 To investigate the possibility of blending of the source, a solution
 without blending parameter, and with blending parameter has been fitted
 to the image subtraction data, for each of the OGLE candidate. To compare
 the Chi-square's to the fit of the unblended and blended model, the F-test
 was performed, in order to estimate  the significance to add a blending
 parameter to the model. The result is presented in table 2. It appears
 immediately that OGLE \#5 and OGLE \#6 are blended microlensing events.
 Considering the flux ratio of the source to the flux of the resolved source,
 it is also very likely that these 2 events result from a large
 magnification of a faint unresolved star. OGLE \#5 had already be identified
 as a lensing of an unresolved star with the previous OGLE data
 (Alard 1997), but the detection of OGLE \#6 as a very blended event is new.
 We emphasize that the confidence of detection the blended nature
 of OGLE \#5 has been greatly improved with the image
 subtraction method. There is also some weak evidence that OGLE \#18 could
 belong to the same category as  OGLE \#5 and  OGLE \#6. 
\subsection{Comparison with the astrometric shift method.}
 In this paragraph we shall make a comparison with the analysis of astrometric
 shifts of the OGLE microlensing candidates performed by Goldberg \& Wozniak
 (1998).
 The two very blended events, OGLE \#5 and  OGLE \#6, both show very
 large astrometric shifts. It confirms the prediction by Han (1998) and
 Goldberg (1998) that lensing
 of very faint highly blended objects should result in the observation of large
 astrometric shifts. On the other hand, some large astrometric shifts 
 can be detected
 without any photometric evidence for them if the separation between 
 the components
 is large and the amplified star is not much fainter than the other
 blend component.
 When the separation between the 
 blend components is
 small and the source is faint compared to the other component, then
 photometric analysis is a good way to detect blending.
 This shows that the two techniques are complementary.
\section{Conclusion.}
 This re-analysis of the OGLE images clearly demonstrates, with overwhelming 
 evidences, the usefulness of the image subtraction method. On the average, we
 noticed an improvement by a factor of 2 in the photometric accuracy in
 comparison to DoPHOT. An improvement of $ \sim 2.5 $ was found for the data
 points taken during the magnification. This makes it possible to refine
 the interpretation of the OGLE microlensing candidates.
 The case of the strongly blended OGLE \#5 is especially important. 
 OGLE \#9 is also
 very interesting, as it shows a light curve that is
 completely different from the DoPHOT light curve. It is a good illustration
 of the large biases that might be present in data processed with DoPHOT.
 However, it was also found that small amplitude
 long term variations were present in the OGLE \#12 and OGLE \#11 
 baseline magnitudes. It is difficult to state if these variations
 make these 2 candidates inconsistent with microlensing. Further
 investigation should be conducted in order to estimate the likelihood of such
 variations in similar stars, which either solar type or subgiants. \\\\
 The case of OGLE \#14 is more problematic since we have very significant
 evidence that the shape of the light magnification differ from the
 point-source microlensing.
 Since we have no data point on the rising branch, it is 
 difficult to prove that this candidate is related to
 cataclysmic variables. It would be valuable to obtain a spectrum
 of the star in order to search for specific emission lines. \\\\
 A refined analysis of the other candidates shows that OGLE \#5 and 
 OGLE \#6 are highly blended microlensing events. They
 result from the magnification of very faint sources which would 
 have been undetectable without the microlensing magnification . \\\\
 It is important to emphasize that the improvement provided
 by the image subtraction method should not be considered only in the case
 of producing individual light curves of some selected objects. This
 method is well suited for global processing of all the microlensing data.
 Due to the fact that the whole linear least-square matrix has to be built
 only once for the reference image, the computing time of the method is very
 reasonable and should compete with DoPHOT. \
 To conclude,
 we would like to emphasize that the improvement that has been found in
 the quality of the light curves is not in any sense specific to OGLE.
 This testing has been conducted with OGLE data, as OGLE has been
 the first to provide massive release of images. 
 However such similar improvements, have already been noticed
 for the light curves of 2 microlensing candidates obtained
 with EROS data (e.g. Afonso {\it et al.}, 1998),
 and it will certainly be seen for other microlensing 
 data-set, as all these projects use DoPHOT or DoPHOT like softwares.
 It is also important to add that some small photometric effects which
 were not found with DoPHOT (see OGLE \# 12 for instance), will probably
 be seen as well when large data-sets from the other microlensing experiments
 will be analyzed with the image subtraction method.\\\\
 The light curves of all OGLE microlensing events can be accessed with
 anonymous ftp at: ftp.iap.fr, directory:   pub/from\_users/alard/ogle.
\begin{table*}[htb]
 \caption[]{Fit of microlensing light curves. We use the following
 notations: D/S is the ratio between DoPHOT residual and the image 
 subtraction residual to the fit of a microlensing light curve. Chi2 
 is the square root of the Chi-square calculated by normalizing
 the residuals to the fit of the image subtraction light curve with
 the Poissonian deviation.
 We present this quantity because it gives directly the percentage 
 of error in addition to the theoretical Poisson error.
 Additionally, the column Mag gives
 the DoPHOT magnitude corresponding to the baseline. 
 The last 2 columns represent the percentage errors for the baseline
 and the magnification periods.
 The capital letter (A) stands for during magnification, and (B) is for 
 data belonging to the baseline of the event. We also present
 the mean * quantity which is the mean of the various parameter calculated
 only for the candidates that are not marked with a *. \\
 Note: For OGLE \#19, Chi2 (A) = 0.92, but we have only a few points during 
 magnification for this microlensing event.
}
  \begin{flushleft}
  \begin{tabular}{llllllll}
   \hline\noalign{\smallskip}
    Candidate  & D/S (B) &  D/S (A) & Chi2 (B) & Chi2 (A) & Mag & \% (B) & \% (A)\\ 
   \hline\noalign{\smallskip}        
             OGLE \#1 & 2.17 & 2.38 & 1.15 & 1.09 & 18.7 & 2.7 & 1.9 \\
             OGLE \#2 & 1.37 & 1.87 & 1.55 & 2.03 & 19.1 & 4.9 & 4.1 \\
             OGLE \#3 & 2.58 & 1.82 & 1.62 & 2.17 & 15.9 & 0.4 & 0.5 \\
             OGLE \#4 & 2.65 & 3.39 & 1.15 & 1.08 & 19.2 & 2.2 & 2.3 \\
             OGLE \#5 & 2.71 & 7.57 & 1.33 & 1.89 & 18.0 & 1.1 & 1.1 \\
             OGLE \#6 & 13.3 & 19.1 & 1.56 & 1.57 & 18.1 & 2.1 & 2.4 \\
             OGLE \#8 & 1.73 & 1.34 & 1.29 & 1.37 & 17.8 & 1.1 & 1.1 \\
             OGLE \#9 & 1.48 & 1.47 & 1.32 & 0.99 & 19.2 & 4.8 & 3.1 \\
             OGLE \#11 & 1.09 & 1.45 & 1.60 & 1.47 & 18.1 & 2.4 & 1.9 \\
             OGLE \#12 & 1.21 & 1.52 & 1.92 & 1.29 & 18.6 & 4.1 & 2.5 \\
             OGLE \#14 & 1.76 & 1.46 & 1.40 & 3.11 & 19.0 & 2.9 & 4.4 \\
             OGLE \#15 & 2.93 & 3.17 & 1.23 & 1.69 & 18.3 & 3.4 & 2.8 \\
             OGLE \#16 & 1.71 & 1.36 & 1.19 & 1.46 & 18.4 & 1.0 & 1.0 \\
             OGLE \#17 & 1.74 & 1.92 & 1.26 & 1.55 & 18.7 & 3.0 & 2.2 \\
             OGLE \#18 & 1.61 & 2.14 & 1.40 & 1.37 & 18.6 & 2.5 & 1.8 \\
             OGLE \#19 & 1.72 & 2.99 & 1.48 & 0.92 & 19.7 & 4.9 & 2.7 \\
             Mean      & 1.96 & 2.5  & 1.4 & 1.56 & 18.5 & 2.7 & 2.2 \\ 
             Mean *    & 2.15 & 2.84 & 1.35 & 1.51 & 18.4 & 2.4 & 1.9\\
    \noalign{\smallskip}
   \hline      
   \end{tabular}
  \end{flushleft}
\end{table*}
\begin{table}[htb]
 \caption[]{Blending analysis. In the first column we present the probability
 that adding a blending parameter to the fitted microlensing model is
 significant. In column 2 we give the flux ratio between the source
 (derived from fitting the blending parameter) and the associated star.
 In the last column we give the magnitude of the source that has been
 calculated by applying the flux ratio correction to the DoPHOT baseline
 magnitude.}
\begin{flushleft}
  \begin{tabular}{llll}
   \hline\noalign{\smallskip}
    Candidate  & Probability & Blending ratio & Mag (I band)\\ 
   \hline\noalign{\smallskip}  
             OGLE \#1 & 0.530 & 1.7 & 21.9 \\
             OGLE \#2 & 0.511 & 1.3 & 19.5 \\
             OGLE \#3 & 0.500 & 1.0 & 15.9 \\
             OGLE \#4 & 0.568 & 1.7 & 19.9 \\
             OGLE \#5 & 1.000 & 49.9 & 22.3 \\
             OGLE \#6 & 0.994 & 16.6 & 21.2 \\
             OGLE \#8 & 0.509 & 1.3 & 18.2 \\
             OGLE \#9 & 0.500 & 1.0 & 19.2 \\
             OGLE \#11 & 0.500 & 1.5 & 18.6 \\
             OGLE \#12 & 0.506 & 2.1 & 19.5 \\
             OGLE \#14 & 0.500 & 1.0 & 19.0 \\
             OGLE \#15 & 0.500 & 1.0 & 18.3 \\
             OGLE \#16 & 0.526 & 5.5 & 20.3 \\
             OGLE \#17 & 0.500 & 1.0 & 18.7 \\
             OGLE \#18 & 0.779 & 16.6 & 21.7 \\
             OGLE \#19 & 0.534 & 2.7 & 20.9 \\
    \noalign{\smallskip}
    \hline      
   \end{tabular}
  \end{flushleft}
\end{table}
\begin{acknowledgements}
 The author would like to thank all the OGLE team for public release
 of the image data-set. Without this unique opportunity, accurate testing
 of the new image subtraction method, and comparison with previous
 method would have been impossible.
 The author is especially indebted to  B. Paczy\'nski 
 for his kind help and comments. 
 It is a pleasure to thank A. Udalski,
  P. Wozniak, E. Aubourg, and D. Goldberg for their comments. 
\end{acknowledgements}
\clearpage
\begin{figure*} 
\centerline{\psfig{angle=90,figure=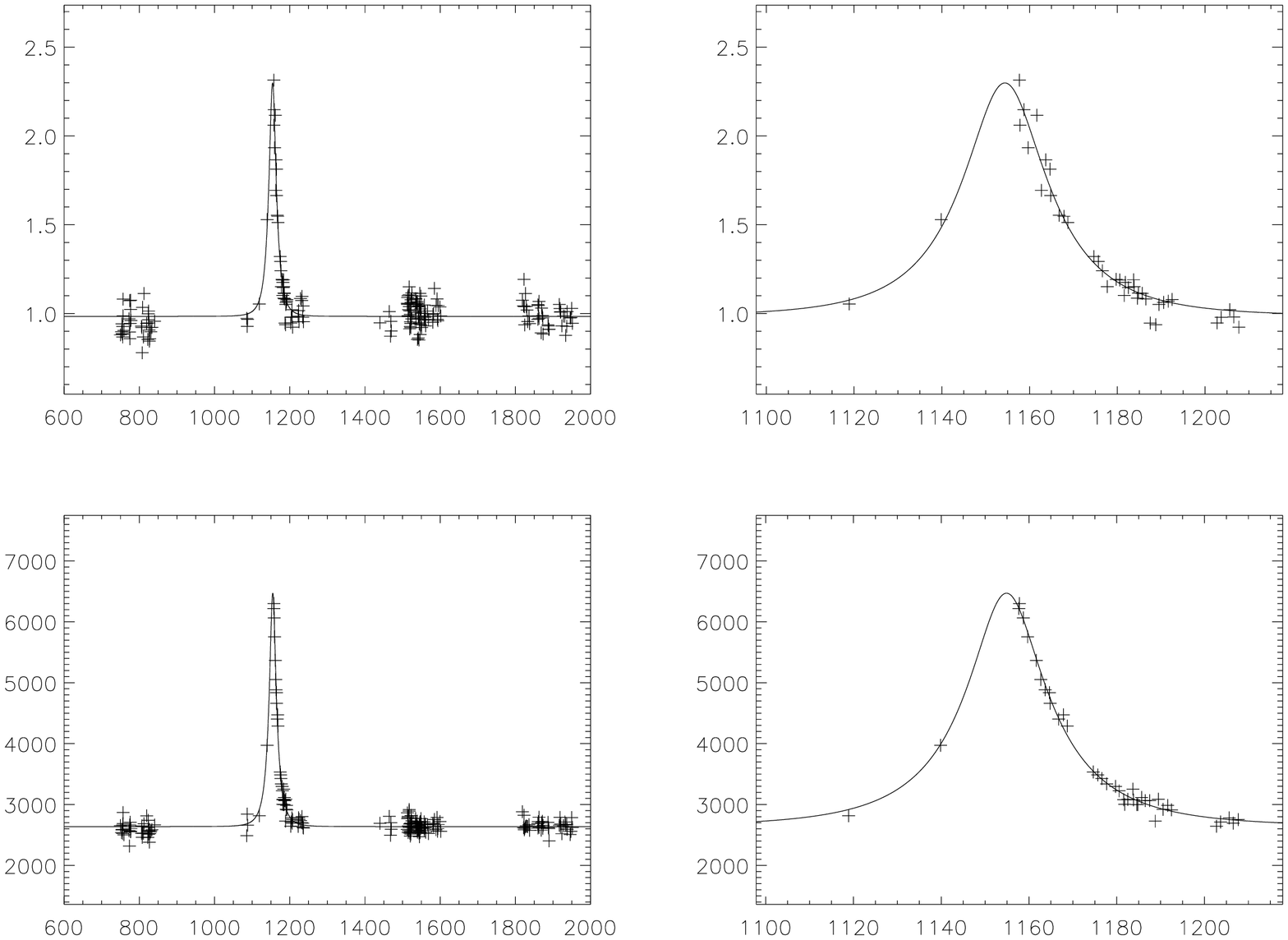,width=20cm}}
\caption{
 OGLE \#1 light curves. 
 The upper light curves were obtained with DoPHOT, the lower
 with the image subtraction method.
 The latter improves the residual by a bit more than a factor of 2.}
\end{figure*}
\begin{figure*} 
\centerline{\psfig{angle=90,figure=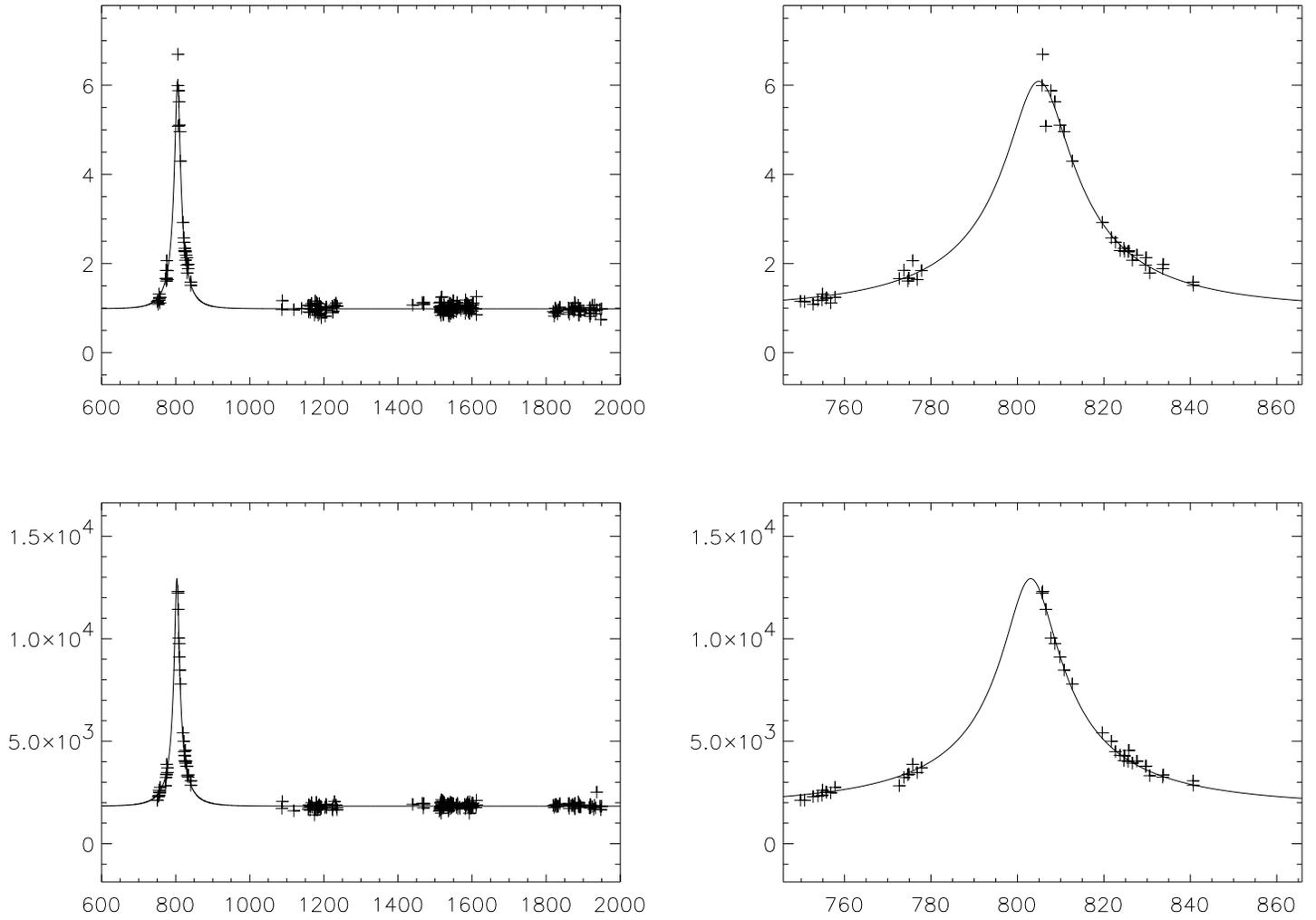,width=20cm}}
\caption{
 OGLE \#2 light curves. 
 The upper light curves were obtained with DoPHOT, the lower
 with the image subtraction method.
 Note the spread for the 
 2 bad seeing points at top of the DoPHOT light curve. The image subtraction
 method reduces considerably the scatter of these bad seeing points.}
\end{figure*}
\begin{figure*} 
\centerline{\psfig{angle=90,figure=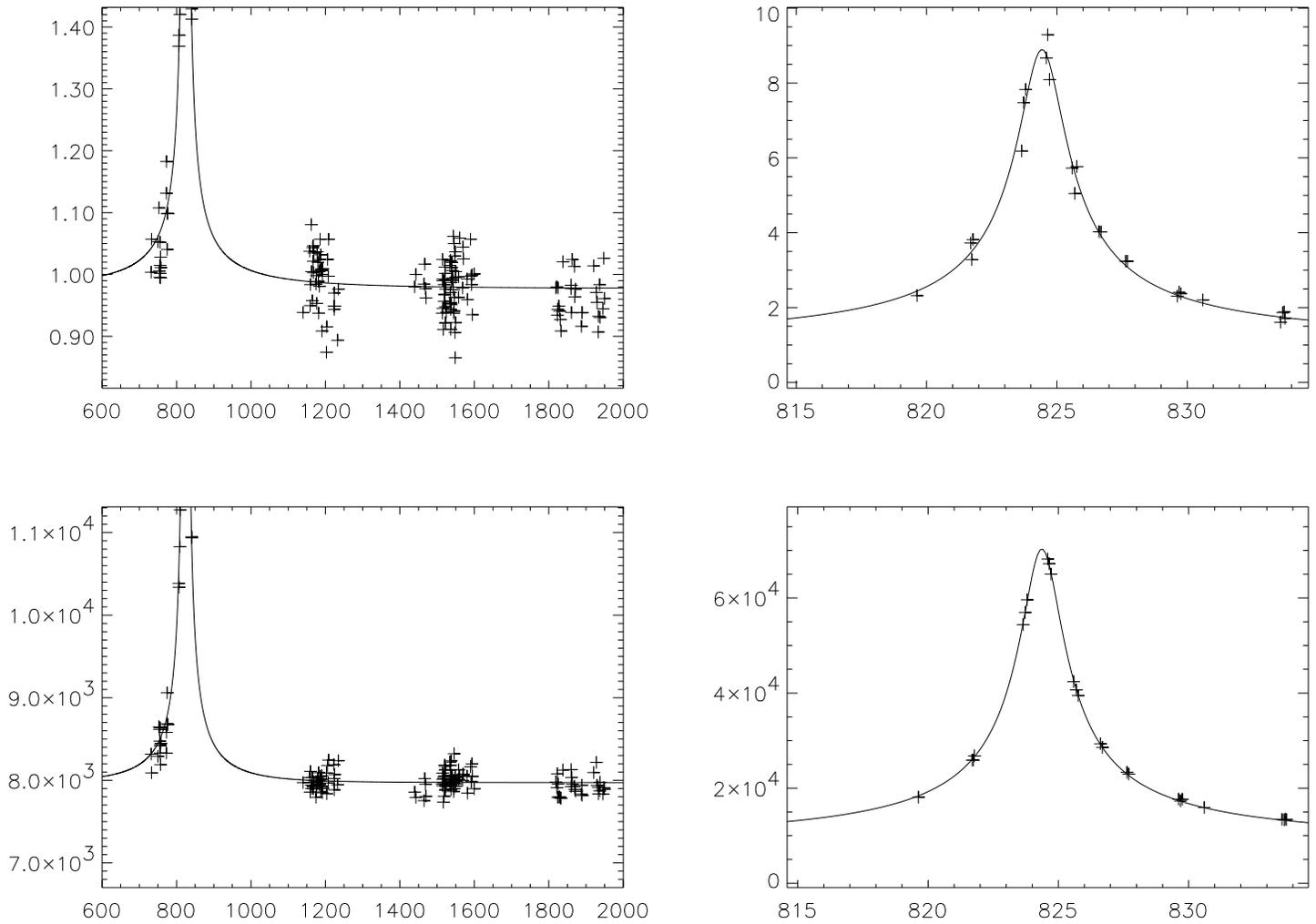,width=20cm}}
\caption{
 OGLE \#5 light curves. 
 The upper light curves were obtained with DoPHOT, the lower
 with the image subtraction method.
 The residuals are improved by more
 than a factor of 7 during the magnification, this is the largest improvement
 obtained by using image subtraction.}
\end{figure*}
\begin{figure*} 
\centerline{\psfig{angle=90,figure=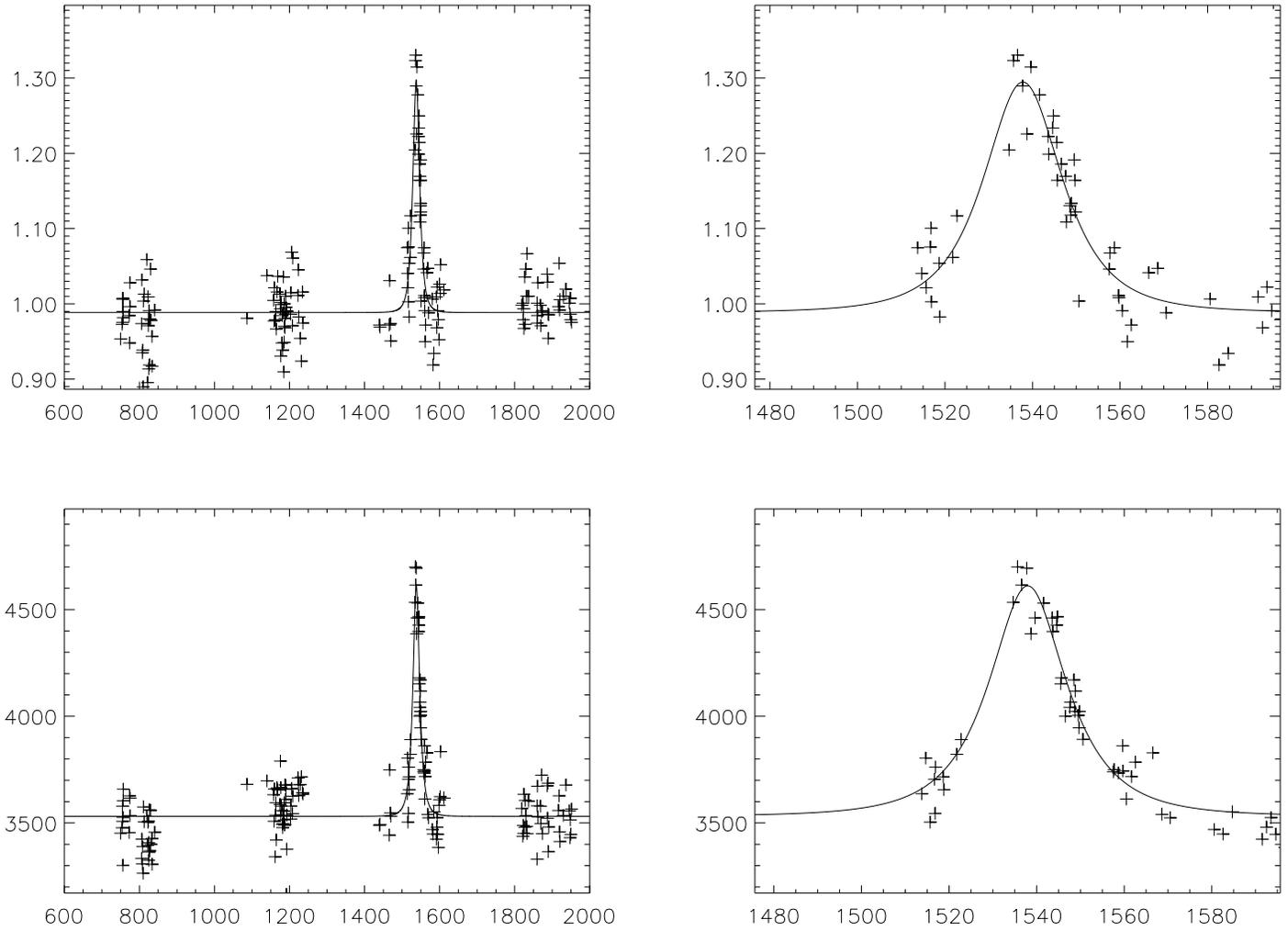,width=20cm}}
\caption{
 OGLE \#11 light curves.
 The upper light curves were obtained with DoPHOT, the lower
 with the image subtraction method.
 Note that in the image 
 subtraction light curve, the clump of points near epoch 800 is slightly
 below the baseline on the mean, and that the clump of points near epoch
 1200 is above the baseline. Due to larger errors, these features are hardly
 visible in the DoPHOT light curve.}
\end{figure*}
\begin{figure*} 
\centerline{\psfig{angle=90,figure=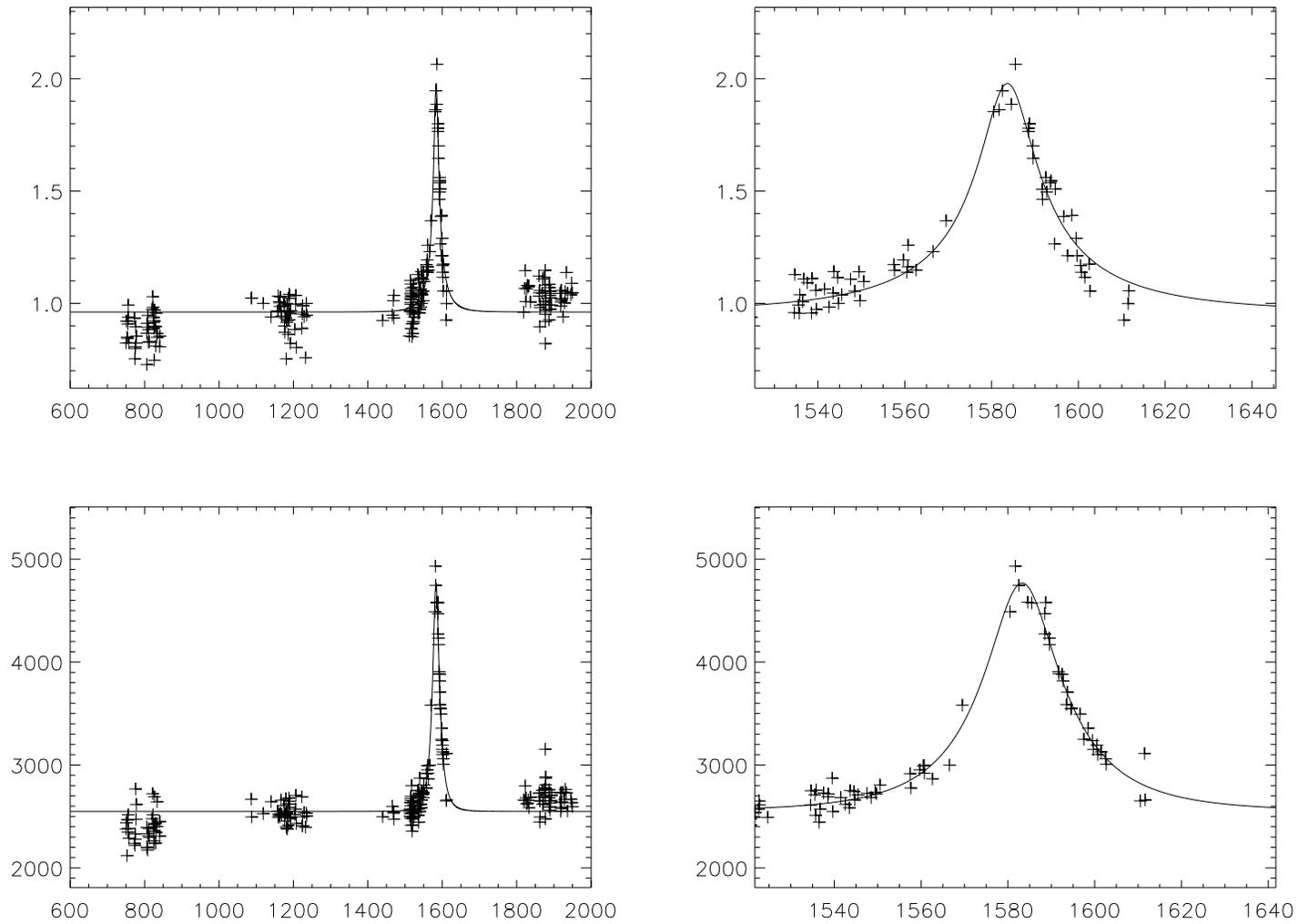,width=20cm}}
\caption{
 OGLE \#12 light curves. 
 The upper light curves were obtained with DoPHOT, the lower
 with the image subtraction method.
 Note the rise of the mean baseline
 magnitude in both, DoPHOT and image subtraction light curves. The difference
 between the 2 edges of the baseline is about 0.2 magnitudes.}
\end{figure*}
\begin{figure*} 
\centerline{\psfig{angle=90,figure=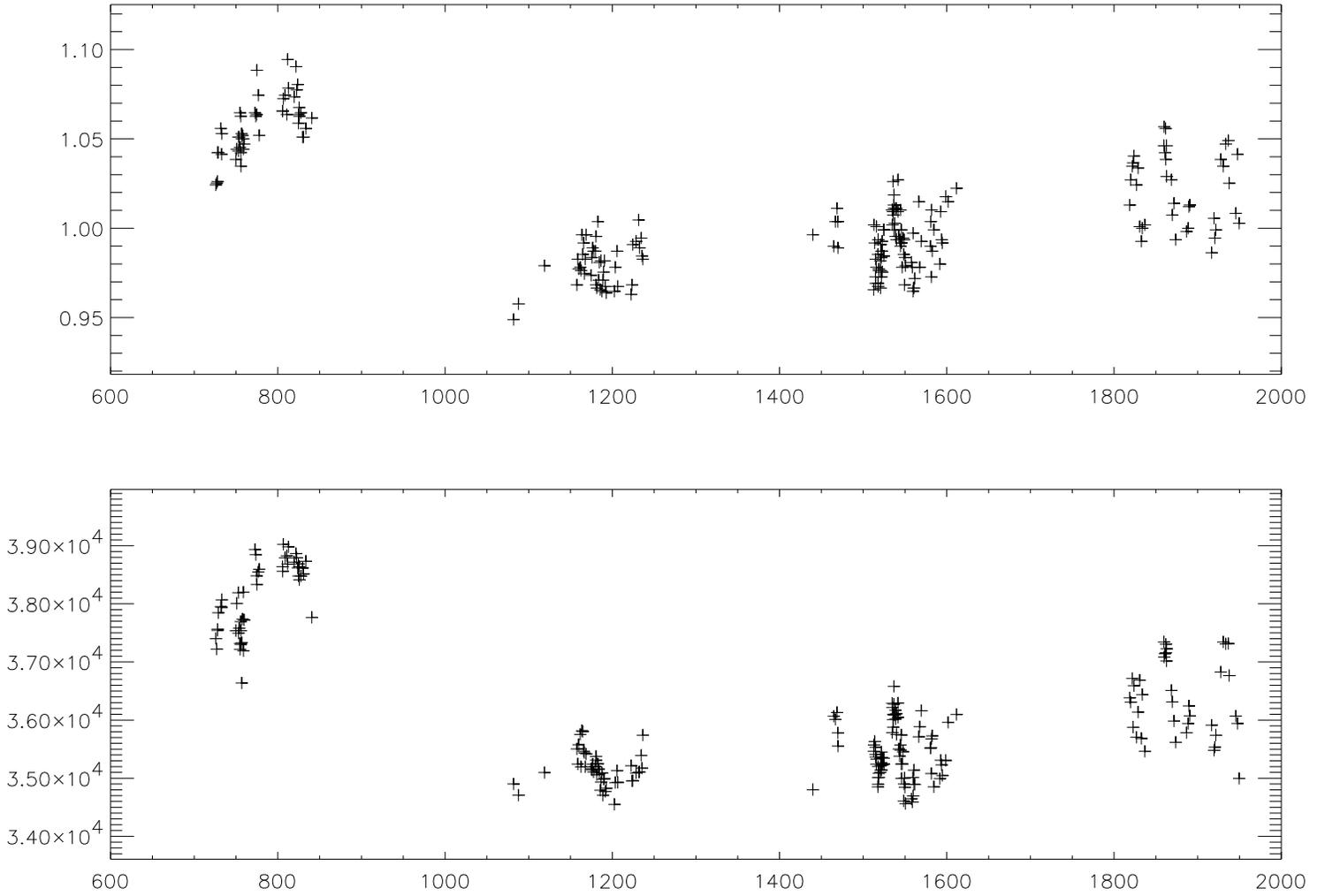,width=20cm}}
\caption{
 OGLE \#10 light curves. Up is DoPHOT, below the light curve obtained
 with the image subtraction method.}
\end{figure*}
\begin{figure*} 
\centerline{\psfig{angle=90,figure=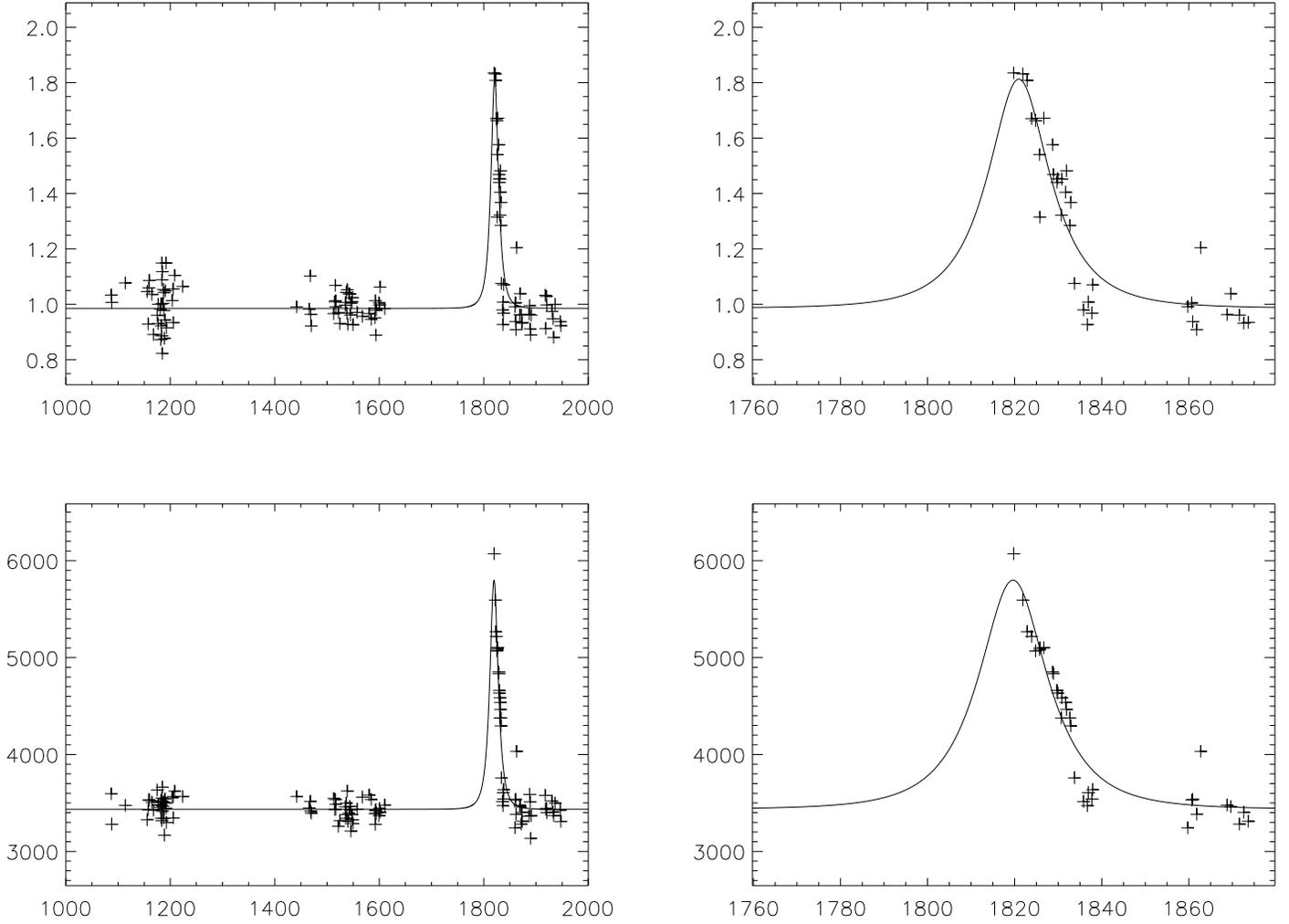,width=20cm}}
\caption{
 OGLE \#14 light curves. 
 The upper light curves were obtained with DoPHOT, the lower
 with the image subtraction method.
 Note the clump of the systematicly deviating
 points at the tip of the decending branch in the image subtraction light curve.
 Also, note the deviating point near epoch 1860 in, both DoPHOT and image
 subtraction light curve.}
\end{figure*}
\begin{figure*} 
\centerline{\psfig{angle=90,figure=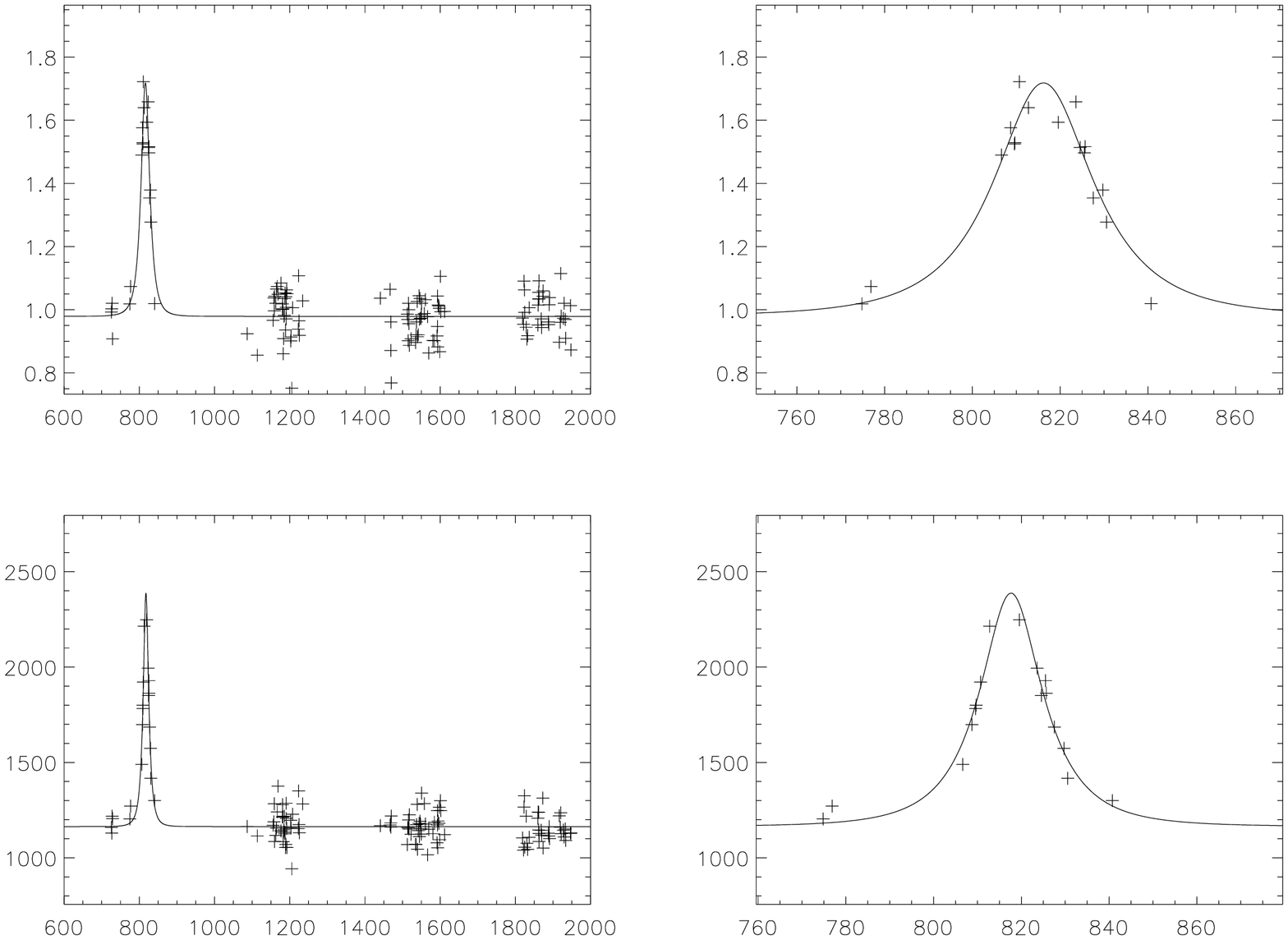,width=20cm}}
\caption{
 OGLE \#9 light curves.
 The upper light curves were obtained with DoPHOT, the lower
 with the image subtraction method.
 Note a completely different shape
 of the light curve obtained with image subtraction. This example
 shows that large biases can be present in the DoPHOT processed data.
 Image subtraction not only improves the photometric accuracy, but it also
 eliminates most biases.}
\end{figure*}

\begin{thebibliography}{}
\bibitem{} Afonso, {\it et al.}, 1998, A\&A, in press (astro-ph/9806380)
\bibitem{} Alard, C., \& Lupton, R.H., 1998, ApJ in Press (astro-ph/9712287)
\bibitem{} Alard, C., 1997, A\&A, 326, p. 1
\bibitem{} Ansari, R., 1996, Vistas in Astronomy, 
 v. 40, p. 519-53
\bibitem{} Bennett, D., 1992, Texas/PASCO, International Conference,
 Berkeley, CA, 13-18 Dec. 1992
\bibitem{} Goldberg, D. M., 1998, ApJ, 498, 156
\bibitem{} Goldberg, D. M., \& Wozniak, P., 1997, AcA, 48, p. 19
 (astro-ph/9712262)
\bibitem{} Han, C., 1997, ApJ, 490, p.51
\bibitem{} Han, C., 1998, astro-ph/9804272
\bibitem{} Schechter, P., Mateo, M., \& Saha, A., 1993, PASP, 105, p. 1342
\bibitem{} Udalski, A., Kubiak, M., \&  Szymanski, M., 1997, AcA, 47, p. 319 
 (astro-ph/9710091)
\bibitem{} Udalski, A., 1993, {\it et al.} AcA, 43, p. 69
\bibitem{} Wozniak, P., \& Szymanski, M., 1998, Astro-ph/9804193
%
%
\end{thebibliography}
\end{document}